\documentclass[american,showkeys,showpacs]{revtex4-1}
\usepackage[T1]{fontenc}
\usepackage[latin9]{inputenc}
\usepackage{amsmath}
\usepackage{amssymb}

\makeatletter

\providecommand{\tabularnewline}{\\}

\@ifundefined{textcolor}{}
{%
 \definecolor{BLACK}{gray}{0}
 \definecolor{WHITE}{gray}{1}
 \definecolor{RED}{rgb}{1,0,0}
 \definecolor{GREEN}{rgb}{0,1,0}
 \definecolor{BLUE}{rgb}{0,0,1}
 \definecolor{CYAN}{cmyk}{1,0,0,0}
 \definecolor{MAGENTA}{cmyk}{0,1,0,0}
 \definecolor{YELLOW}{cmyk}{0,0,1,0}
}

\usepackage{url}
\usepackage{graphicx}
\usepackage{tikz}
\usepackage{tikz-3dplot}

\makeatother

\usepackage{babel}
\begin{document}

\title{Qubits, Weyl spinors, quantum NOT gates, and dynamical decoupling}

\author{R. Romero}

\email{romero@ciencias.unam.mx}

\selectlanguage{american}%

\affiliation{Instituto de F\'isica, Universidad Nacional Aut\'onoma de M\'exico, Apartado
Postal 20-364, M\'exico 01000, D. F., M\'exico}
\begin{abstract}
An equivalence is established between orthogonal pure state qubits
on the Bloch sphere and massless Weyl spinors, when the Bloch vector
is taken as the physical three-momentum. A family of unitary, coordinate
dependent transformations is obtained which connects orthogonal combinations
of the basis states of a two-level quantum system. It is shown that
a subset of these transformations possesses the novel feature of effecting
a point inversion by means of a rotation. For qubits, these transformations
act as quantum NOT/parity gates, and also as flipping operators that
exactly cancel decoherence in a dynamical decoupling setting. For
Weyl spinors they provide, at the relativistic quantum level, a unitary
symmetry transformation for the Weyl equations.
\end{abstract}



\maketitle

\section{Introduction}

In quantum information theory the analogue of a classical bit, a qubit,
is provided by a general linear complex combination of the orthogonal
states of a two-level system. The orthogonal states $\left|0\right\rangle $
and $\left|1\right\rangle $ of the two-level system constitute the
computational basis. If the computational basis is assigned to the
north and south poles of a unit sphere, known as the Bloch sphere
(figure 1), any point on the surface of the sphere represents a pure
state qubit\cite{nielsen2010quantum,barnett2009quantum}, with the
corresponding orthogonal state given by the point diametrically opposed
to it: the antipodal point\cite{williams2010explorations}. 

A quantum gate is a unitary matrix acting on one or more qubits, and
it is the basic building block for quantum circuits. Classically,
the negation or NOT gate swaps the logic binaries 0 and 1. In quantum
computing, a quantum NOT gate would, ideally, transform an arbitrary
qubit, including the computational basis, into its orthogonal state,
mimicking the action of its classical counterpart, and thus providing
a universal quantum NOT gate, where in this context universality is
to be understood as the ability to output the orthogonal state to
any given input%
\footnote{This definition of universality should not be confused with the standard
definition, whereby a set of operations can be reduced to the action
of a gate%
}. However, it is known that no \emph{fixed }unitary matrix exists
for that purpose\cite{PhysRevA.60.R2626,Buifmmodecheckzelsevzfiek2000,williams2010explorations,barnett2009quantum}. 

A quantum NOT gate is analogous to a spin flip operator\cite{Gisin:1999sp},
and because of the antipodal character of pure state qubits, also
to a parity operator. In fact, the universal quantum NOT gate, as
described above, can be characterized as a point inversion through
the origin of the Bloch sphere\cite{PhysRevA.60.R2626}. Flipping
operators find important applications in dynamical decoupling schemes,
where they are used to reverse decoherence along a given axis of the
Bloch sphere. This is done by applying sequences of pulses that, on
average, transform the state to its mirror state across the relevant
symmetry plane\cite{raey}. The pulses must anti-commute with the
effective interaction Hamiltonian of system plus bath, and thus an
operator that anti-commutes with a general interaction Hamiltonian
would then constitute a universal dynamical decoupler.

On a seemingly unrelated topic, it is well known that the free Dirac
equation decouples in the massless limit into the Weyl equations\cite{Pal:2010ih,Muller-Kirsten2010,pokorski2000gauge},
whose solutions are two-component spinors of definite helicity, which
is the spin projection along the direction of the physical momentum:
left-handed spinors for helicity $-1$, and right-handed spinors for
helicity $+1$. These classical, \emph{c}-number spinors, are also
regarded as twistors\cite{srednicki2007quantum}. Left and right-handed
spinors belong to nonequivalent irreducible representations of the
Lorentz group\cite{tung1985group,s1995group,ryder1996quantum}, connected
by parity. They are also related by an anti-unitary transformation,
the so called Wigner transformation or spin flip operation\cite{wigner1959group,peskin1995introduction,ryder1996quantum}.
The Wigner transformation is a symmetry transformation of the Weyl
equations, analogous to parity\cite{PhysRev.107.307,PhysRev.106.821,Sakurai1964}.
As in the qubit case, there is no fixed unitary matrix that connects
Weyl spinors of different helicity.

The purpose of this letter is two-fold: one is to present an analogy
between pure state qubits and massless Weyl spinors, thus providing
a connection between previously thought unrelated topics, and the
second is to present a family of unitary transformations connecting
the orthogonal combinations of the basis states of a two-level system,
and to show their applications to qubits and Weyl spinors. These transformations
depend on the states and hence are continuous, coordinate dependent
transformations, and in one instance they have the remarkable and
novel property of effecting a point inversion by means of a rotation.
For qubits, the transformations provide a family of NOT/parity gates
which anti-commute with a general interaction Hamiltonian of system
plus bath, and thus exactly cancel decoherence in a dynamical decoupling
setting. Also, a one parameter transformation is obtained that satisfy
the criterion for universality described above. For Weyl spinors,
the transformations comprise a novel symmetry transformation for the
Weyl equations.

The organization is as follows: section II provides the relation between
qubits and massless Weyl spinors, and it is also pointed out that
there are two types of spinors with different transformation phases
under a discrete parity transformation $\mathcal{P}$. Section III
introduces the unitary transformations and their properties in a general
setting. Section IV deals with the application of the transformations
to qubits, and shows their role in dynamical decoupling. Section V
presents the new symmetry transformation of the Weyl equations, and
also shows the role of the transformations in the definition of twistors.
Finally, concluding remarks are given. Natural units with $\hbar=c=1$
are used throughout.

\section{Qubits as massless Weyl spinors}

In regards to the unit Bloch sphere \cite{nielsen2010quantum,barnett2009quantum},
a general qubit is given by

\begin{equation}
\left|\chi_{+}\right\rangle =\cos\left(\frac{\theta}{2}\right)\left|0\right\rangle +e^{i\varphi}\sin\left(\frac{\theta}{2}\right)\left|1\right\rangle ,\label{eq:1}
\end{equation}

\noindent where $\theta$ and $\varphi$ are the spherical polar angles
and $\{\left|0\right\rangle ,\left|1\right\rangle \}$ is a suitable
basis of the two-level system, respectively represented as the north
and south poles of the sphere (Figure 1).
\begin{figure}
\begin{centering}
\tdplotsetmaincoords{70}{110}

\pgfmathsetmacro{\rvec}{1.3} 
\pgfmathsetmacro{\thetavec}{45} 
\pgfmathsetmacro{\phivec}{45}

\pgfmathsetmacro{\thetavecp}{135} 
\pgfmathsetmacro{\phivecp}{225}

\begin{tikzpicture}[scale=3,tdplot_main_coords] 

\draw (2.9,0,0.9) arc (180:360:0.99cm and 0.3cm);     
\draw[dashed] (2.9,0,0.9) arc (180:0:0.99cm and 0.3cm);          
\draw (0,0) circle (1cm);     

\draw[thick,->] (0,0,0) -- (1.09,0,0) node[anchor=south east]{$p_{x}$}; 
\draw[thick,->] (0,0,0) -- (0,1.05,0) node[anchor=north west]{$p_{y}$}; 
\draw[thick,->] (0,0,0) -- (0,0,1.05) node[anchor=north west]{$p_{z}$};
\draw[dashed] (0,0,0) -- (0,0,-1.05);
\node at (0,0,-1.15) {$\left|1\right>$};
\node at (0,0,1.15) {$\left|0\right>$};

\coordinate (O) at (0,0,0);

\tdplotsetcoord{P}{\rvec}{\thetavec}{\phivec}
\tdplotsetcoord{-P}{\rvec}{\thetavecp}{\phivecp}     
\draw[thick,->] (O) -- (P) node[above right] {$\left|\chi_{+}\right>$};  
\draw[thick,dashed,->] (O) -- (-P) node[below] {$\left|\chi_{-}\right>$};  
\draw[dashed] (O) -- (Pxy);     
\draw[dashed] (P) -- (Pxy);     
\tdplotdrawarc{(O)}{0.3}{0}{\phivec}{anchor=north west}{$\varphi$}     
\tdplotsetthetaplanecoords{\phivec}     
\tdplotdrawarc[tdplot_rotated_coords]{(0,0,0)}{0.4}{0}%         
{\thetavec}{anchor=south west}{$\theta$}

\end{tikzpicture}
\par\end{centering}

\protect\caption{Unit Bloch Sphere. The computational basis is mapped to the north
and south poles of the sphere. The orthogonal pure states $\left|\chi_{+}\right\rangle $
and $\left|\chi_{-}\right\rangle $ are antipodal and they can be
described in terms of the polar angles $\left(\theta,\varphi\right)$
and the computational basis, as in Eqs. (\ref{eq:1}) and (\ref{eq:5}).}

\end{figure}
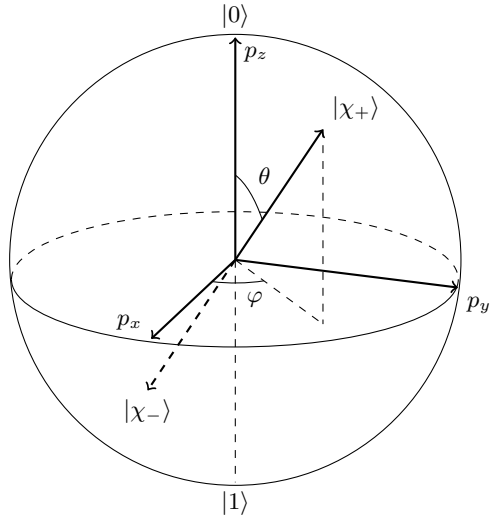
 The subscript in $\left|\chi_{+}\right\rangle $ denotes the helicity,
as will be explained shortly. For a spin $1/2$ system, and using
the Pauli matrices

\begin{equation}
\begin{array}{ccc}
\sigma_{1}=\left(\begin{array}{cc}
0 & 1\\
1 & 0
\end{array}\right), & \sigma_{2}=\left(\begin{array}{cc}
0 & -i\\
i & 0
\end{array}\right), & \sigma_{3}=\left(\begin{array}{cc}
1 & 0\\
0 & -1
\end{array}\right),\end{array}\label{eq:2}
\end{equation}

\noindent the computational basis states are the eigenstates of $\sigma_{3}$

\begin{equation}
\begin{array}{cc}
\left|0\right\rangle =\left(\begin{array}{c}
1\\
0
\end{array}\right), & \left|1\right\rangle =\left(\begin{array}{c}
0\\
1
\end{array}\right).\end{array}\label{eq:3}
\end{equation}

\noindent Hence, the qubit in Eq. (\ref{eq:1}) can be rewritten as 

\begin{equation}
\left|\chi_{+}\right\rangle =\begin{pmatrix}\cos\left(\frac{\theta}{2}\right)\\
e^{i\varphi}\sin\left(\frac{\theta}{2}\right)
\end{pmatrix}.\label{eq:4}
\end{equation}

\noindent Points on the surface of the Bloch sphere connected by a
diameter correspond to orthogonal pure states, and the antipodal state
to $\left|\chi_{+}\right\rangle $ is given by 

\begin{equation}
\left|\chi_{-}\right\rangle =-e^{-i\varphi}\sin\left(\frac{\theta}{2}\right)\left|0\right\rangle +\cos\left(\frac{\theta}{2}\right)\left|1\right\rangle ,\label{eq:5}
\end{equation}

\noindent or equivalently

\begin{equation}
\left|\chi_{-}\right\rangle =\begin{pmatrix}-e^{-i\varphi}\sin\left(\frac{\theta}{2}\right)\\
\cos\left(\frac{\theta}{2}\right)
\end{pmatrix}.\label{eq:6}
\end{equation}

\noindent The qubits $\left|\chi_{+}\right\rangle $ and $\left|\chi_{-}\right\rangle $
form an orthonormal set.

Let us now consider the free, massless Dirac equation 

\begin{equation}
i\gamma^{\mu}\partial_{\mu}\Psi=0,\label{eq:7}
\end{equation}

\noindent where the gamma matrices $\gamma^{\mu}=\left(\gamma^{0},\boldsymbol{\gamma}\right)$
satisfy the Clifford algebra relation $\gamma^{\mu}\gamma^{\nu}+\gamma^{\nu}\gamma^{\mu}=2g^{\mu\nu}$,
with $g^{\mu\nu}$ the metric tensor with signature $\text{diag}(1,-1,-1,-1)$.
Using the Weyl representation of the gamma matrices 
\begin{equation}
\begin{array}{c}
\gamma^{0}=\left(\begin{array}{rr}
0 & 1\\
1 & 0
\end{array}\right),\,\boldsymbol{\gamma}=\left(\begin{array}{rr}
0 & \boldsymbol{\sigma}\\
-\boldsymbol{\sigma} & 0
\end{array}\right),\end{array}\label{eq:8}
\end{equation}

\noindent here denoted in $2\times2$ block form, with $\boldsymbol{\sigma}=\left(\sigma_{1},\sigma_{2},\sigma_{3}\right)$,
Eq. (\ref{eq:7}) decouples into the two Weyl equations\cite{Pal:2010ih,Muller-Kirsten2010,pokorski2000gauge}

\begin{equation}
i\frac{\partial}{\partial t}\psi_{+}=-i\boldsymbol{\nabla}\cdot\boldsymbol{\sigma}\psi_{+},\label{eq:9}
\end{equation}

\begin{equation}
i\frac{\partial}{\partial t}\psi_{-}=i\boldsymbol{\nabla}\cdot\boldsymbol{\sigma}\psi_{-}.\label{eq:10}
\end{equation}

\noindent Inserting the plane wave solutions

\begin{equation}
\psi_{\pm}=\chi_{\pm}(\mathbf{p})\exp\left\{ -i\left(Et-\mathbf{x}\cdot\mathbf{p}\right)\right\} ,\label{eq:11}
\end{equation}

\noindent into Eqs. (\ref{eq:9}) and (\ref{eq:10}), respectively,
we get

\begin{equation}
\boldsymbol{\sigma}\cdot\mathbf{\hat{p}}\,\chi_{\pm}(\mathbf{p})=\pm\chi_{\pm}(\mathbf{p}),\label{eq:12}
\end{equation}

\noindent with $\mathbf{\hat{p}}=\mathbf{p}/\left|\mathbf{p}\right|$
and $p^{0}=E=\left|\mathbf{p}\right|$. Thus, the spinors $\chi_{\pm}(\mathbf{p})$
are helicity eigenstates, with the sign labeling the helicity. Representing
the three-momentum in spherical polar coordinates

\begin{equation}
\mathbf{\hat{p}}=\left(\sin\theta\cos\varphi,\sin\theta\sin\varphi,\cos\theta\right),\label{eq:13}
\end{equation}

\noindent the following solutions to Eq. (\ref{eq:12}) are obtained

\begin{gather}
\begin{gathered}\chi_{+}(\mathbf{p})=\begin{pmatrix}\cos\left(\frac{\theta}{2}\right)\\
e^{i\varphi}\sin\left(\frac{\theta}{2}\right)
\end{pmatrix},\\
\chi_{-}(\mathbf{p})=\begin{pmatrix}-e^{-i\varphi}\sin\left(\frac{\theta}{2}\right)\\
\cos\left(\frac{\theta}{2}\right)
\end{pmatrix}.
\end{gathered}
\label{eq:14}
\end{gather}

\noindent which are just Eqs. (\ref{eq:4}) and (\ref{eq:6}), hence
the subscripts in the latter equations. The relation between massless
Weyl spinors and orthogonal pure state qubits is now established:
if the Bloch vector represents the physical three-momentum $\mathbf{\hat{p}}$,
any two pair of antipodal pure state qubits are just helicity eigenstates,
which in turn are momentum space solutions to the Weyl equations. 

The spinors in Eq. (\ref{eq:14}) are not unique, since any other
pair that differs by an overall phase is also a solution to Eq. (\ref{eq:12}).
One such other pair of orthonormal helicity spinors, often found in
the literature, is given by\cite{berestetskii1982quantum,sakurai2014modern}

\begin{gather}
\begin{split}\eta_{+}(\mathbf{p})=\begin{pmatrix}e^{-i\varphi/2}\cos\left(\frac{\theta}{2}\right)\\
e^{i\varphi/2}\sin\left(\frac{\theta}{2}\right)
\end{pmatrix}\\
\eta_{-}(\mathbf{p})=\begin{pmatrix}-e^{-i\varphi/2}\sin\left(\frac{\theta}{2}\right)\\
e^{i\varphi/2}\cos\left(\frac{\theta}{2}\right)
\end{pmatrix}
\end{split}
\label{eq:15}
\end{gather}

\noindent which correspond to the qubits\cite{gerry2005introductory,bellac2006short}

\begin{equation}
\left|\eta_{+}\right\rangle =e^{-i\varphi/2}\cos\left(\frac{\theta}{2}\right)\left|0\right\rangle +e^{i\varphi/2}\sin\left(\frac{\theta}{2}\right)\left|1\right\rangle ,\label{eq:16}
\end{equation}

\begin{equation}
\left|\eta_{-}\right\rangle =-e^{-i\varphi/2}\sin\left(\frac{\theta}{2}\right)\left|0\right\rangle +e^{i\varphi/2}\cos\left(\frac{\theta}{2}\right)\left|1\right\rangle .\label{eq:17}
\end{equation}

\noindent These qubits are related to $\left|\chi_{\pm}\right\rangle $
by a phase

\begin{equation}
\left|\chi_{\pm}\right\rangle =e^{\pm i\varphi/2}\left|\eta_{\pm}\right\rangle ,\label{eq:18}
\end{equation}

\noindent and hence we would not expect, in principle, any physical
difference with respect to $\left|\chi_{\pm}\right\rangle $. However,
they transform differently under two successive parity transformations\cite{PhysRev.79.495,berestetskii1982quantum,Berg:2000ne}.
To show it, let us implement twice the standard discrete parity transformation
$\mathcal{P}$

\noindent 
\begin{equation}
\mathcal{P}:\left(\theta,\varphi\right)\rightarrow\left(\pi-\theta,\phi+\pi\right),\label{eq:18-1}
\end{equation}

\noindent on the four qubit states. The results are

\begin{equation}
\left|\chi_{\pm}\right\rangle \xrightarrow{\mathcal{P}}\mp e^{\pm i\varphi}\left|\chi_{\mp}\right\rangle \xrightarrow{\mathcal{P}}\left|\chi_{\pm}\right\rangle ,\label{eq:19}
\end{equation}

\begin{equation}
\left|\eta_{\pm}\right\rangle \xrightarrow{\mathcal{P}}i\left|\eta_{\mp}\right\rangle \xrightarrow{\mathcal{P}}-\left|\eta_{\pm}\right\rangle .\label{eq:20}
\end{equation}

\noindent Thus, $\mathcal{P}^{2}=-1$ for qubits $\left|\eta_{\pm}\right\rangle $,
while $\mathcal{P}^{2}=1$ for $\left|\chi_{\pm}\right\rangle $.%
\footnote{For classical, \emph{c}-number spinors, it is known that the ones
with $\mathcal{P}^{2}=-1$ describe Majorana spinors (see references
\cite{berestetskii1982quantum} and \cite{Berg:2000ne}), however,
there is no evidence that they make any difference for quantized fields.%
} The density matrices are just helicity projection operators, and
they do not distinguish between states of different $\mathcal{P}^{2}$

\begin{gather}
\begin{split}\rho_{+}=\left|\chi_{+}\right\rangle \left\langle \chi_{+}\right|=\left|\eta_{+}\right\rangle \left\langle \eta_{+}\right|=\frac{1}{2}\left(I_{2}+\boldsymbol{\sigma}\cdot\mathbf{\hat{p}}\right),\\
\rho_{-}=\left|\chi_{-}\right\rangle \left\langle \chi_{-}\right|=\left|\eta_{-}\right\rangle \left\langle \eta_{-}\right|=\frac{1}{2}\left(I_{2}-\boldsymbol{\sigma}\cdot\mathbf{\hat{p}}\right),
\end{split}
\label{eq:20-1}
\end{gather}

\noindent where $I_{2}$ is the $2\times2$ identity matrix.

\section{Two level-system unitary transformations}

As previously stated, it is not possible to change the general qubit
$\left|\chi_{+}\right\rangle $ ($\left|\eta_{+}\right\rangle $)
into the orthogonal state $\left|\chi_{-}\right\rangle $ ($\left|\eta_{-}\right\rangle $)
with a sngle unitary \emph{constant} matrix, but this can be done
with an anti-unitary transformation \cite{williams2010explorations,2005quant.ph.12172C,barnett2009quantum}.
For Weyl spinors this is known as the Wigner transformation\cite{wigner1959group,peskin1995introduction,ryder1996quantum}

\begin{align}
\begin{split}-i\sigma_{2}\chi_{+}^{*} & =+\chi_{-},\\
-i\sigma_{2}\chi_{-}^{*} & =-\chi_{+},
\end{split}
\label{eq:3.1}
\end{align}

\noindent which is anti-unitary because of complex conjugation. For
qubits, the anti-unitary character of the Wigner transformation means
that it cannot be realized as a quantum gate.

These results notwithstanding, it is well known that $\text{SU}(2)$
provides a double cover of $\text{SO}(3)$ because the latter is not
simply connected. This means that there are closed loops in the $\text{SO}(3)$
sphere, the group manifold, which cannot be continuously reduced to
a point, with the archetypical example being the closed loop which
connects two antipodal points on the surface of the sphere, just as
in the case of orthogonal pure state qubits. Given that $\text{SO}(3)$
is the three-dimensional rotation group, and that is isomorphic to
the group of unitary transformations up to a phase, the so called
projective representation group, one would expect that a unitary \emph{continuous}
transformation connecting the qubits can actually be obtained. The
$\text{SO}(3)$ manifold is $S^{3}$, a four dimensional unit sphere,
but we can obtain a relation to the standard three-dimensional unit
sphere $S^{2}$ (the Bloch sphere) by means of the quotient group
$\text{SO}(3)/\text{SO}(2)$\cite{Mukunda2003,d2007introduction},
then we have the following equivalence

\begin{equation}
\text{SU}(2)/\text{U}(1)\simeq\text{SO}(3)/\text{SO}(2)\simeq\mathbb{C}\text{P}^{1}\simeq S^{2},\label{eq:3.2}
\end{equation}

\noindent where $\mathbb{C}\text{P}^{1}$ is the complex projective
line, which is just the Hilbert space of a two-level quantum system.
Equation (\ref{eq:3.2}) provides the group theoretical basis for
the existence of unitary transformations connecting orthogonal pure
state qubits, although a rigorous mathematical justification requires
further analysis. Regardless, the transformations can be explicitly
given. In this section I present a general realization of such transformations,
and specific examples for qubits/Weyl spinors will be given in the
following section.

Let us consider the following orthonormal states

\begin{gather}
\begin{gathered}\left|\Psi\right\rangle =\alpha\left|0\right\rangle +\beta\left|1\right\rangle ,\\
\left|\Psi_{\perp}\right\rangle =-\beta^{*}\left|0\right\rangle +\alpha^{*}\left|1\right\rangle ,
\end{gathered}
\label{eq:3.3}
\end{gather}

\noindent where $\left|\alpha\right|^{2}+\left|\beta\right|^{2}=1$,
and $\{\left|0\right\rangle ,\left|1\right\rangle \}$ is the basis
of a two-level system. From these states, the following matrix is
obtained

\begin{equation}
\Pi=\delta_{1}\left|\Psi\right\rangle \left\langle \Psi_{\perp}\right|+\delta_{2}\left|\Psi_{\perp}\right\rangle \left\langle \Psi\right|,\label{eq:3.4}
\end{equation}

\noindent with $\delta_{1}$ and $\delta_{2}$ arbitrary phases. In
the basis of Eq. (\ref{eq:3.3}) the explicit matrix form is

\begin{equation}
\Pi=\begin{pmatrix}\left\langle \Psi\right|\Pi\left|\Psi\right\rangle  & \left\langle \Psi\right|\Pi\left|\Psi_{\perp}\right\rangle \\
\left\langle \Psi_{\perp}\right|\Pi\left|\Psi\right\rangle  & \left\langle \Psi_{\perp}\right|\Pi\left|\Psi_{\perp}\right\rangle 
\end{pmatrix}=\begin{pmatrix}0 & \delta_{1}\\
\delta_{2} & 0
\end{pmatrix}.\label{eq:3.5}
\end{equation}

\noindent Hence, 

\begin{equation}
\det\Pi=-\delta_{1}\delta_{2}.\label{eq:3.6}
\end{equation}

\noindent Taking the Hermitian conjugate of $\Pi$ gives

\begin{equation}
\Pi^{\dagger}=\delta_{2}^{*}\left|\Psi\right\rangle \left\langle \Psi_{\perp}\right|+\delta_{1}^{*}\left|\Psi_{\perp}\right\rangle \left\langle \Psi\right|,\label{eq:3.7}
\end{equation}

\noindent and so

\begin{equation}
\Pi\Pi^{\dagger}=\Pi^{\dagger}\Pi=\begin{pmatrix}\left|\delta_{1}\right|^{2} & 0\\
0 & \left|\delta_{2}\right|^{2}
\end{pmatrix}.\label{eq:3.8}
\end{equation}

\noindent The action of $\Pi$ on the states in Eq. (\ref{eq:3.3})
is easily obtained

\begin{align}
\begin{split}\Pi\left|\Psi\right\rangle =\delta_{2}\left|\Psi_{\perp}\right\rangle ,\\
\Pi\left|\Psi_{\perp}\right\rangle =\delta_{1}\left|\Psi\right\rangle .
\end{split}
\label{eq:3.9}
\end{align}

\noindent If we now make the choices

\begin{gather}
\begin{gathered}\delta_{1}\delta_{2}=-1,\\
\left|\delta_{1}\right|^{2}=\left|\delta_{2}\right|^{2}=1,
\end{gathered}
\label{eq:3.10}
\end{gather}

\noindent then from Eqs. (\ref{eq:3.6}) and (\ref{eq:3.8}) we have
that $\Pi$ is a unitary matrix with unit determinant, and thus is
a pure rotation belonging to $\text{SU}(2)$. Moreover, because of
the antipodal character of the orthogonal states and the results in
Eq. (\ref{eq:3.9}), we obtain the remarkable result that $\Pi$ produces
a point inversion through the origin of the Bloch sphere by means
of a rotation. It must be emphasized that this result does not contradict
the assertion at the beginning of this section, because $\Pi$ is
coordinate dependent by construction, so rather than having a single
constant matrix we obtain a family of matrices, one for each pair
of orthogonal pure states, which transforms the states into each other
up to phases, and by imposing the conditions in Eq. (\ref{eq:3.10})
a subset of that family is made up of parity changing rotations.

\section{Applications to qubits}

\subsection{NOT/parity quantum gates}

The results of the previous section can now be readily applied to
the qubits $\left|\chi_{\pm}\right\rangle $ and $\left|\eta_{\pm}\right\rangle $.
Let us first consider the matrices

\begin{equation}
\begin{array}{l}
P_{1}=e^{-i\varphi}\left|\chi_{+}\right\rangle \left\langle \chi_{-}\right|-e^{i\varphi}\left|\chi_{-}\right\rangle \left\langle \chi_{+}\right|,\\
P_{2}=i\left|\eta_{+}\right\rangle \left\langle \eta_{-}\right|+i\left|\eta_{-}\right\rangle \left\langle \eta_{+}\right|,
\end{array}\label{eq:4.1}
\end{equation}

\noindent which are unitary and of unit determinant as can be checked
with the aid of Eq. (\ref{eq:3.10}). In matrix form they are given
by

\begin{equation}
P_{1}=\exp\left(i\frac{\pi}{2}\mathbf{a}\cdot\boldsymbol{\sigma}\right)=\begin{pmatrix}0 & e^{-i\varphi}\\
-e^{i\varphi} & 0
\end{pmatrix},\label{eq:23}
\end{equation}

\begin{equation}
P_{2}=\exp\left(i\frac{\pi}{2}\mathbf{b}\cdot\boldsymbol{\sigma}\right)=\begin{pmatrix}-i\sin\theta & ie^{-i\varphi}\cos\theta\\
ie^{i\varphi}\cos\theta & i\sin\theta
\end{pmatrix},\label{eq:24}
\end{equation}

\noindent where the vectors

\begin{gather}
\begin{gathered}\mathbf{a}=\left(-\sin\varphi,\cos\varphi,0\right),\\
\mathbf{b}=\left(\cos\theta\cos\varphi,\cos\theta\sin\varphi,-\sin\theta\right),
\end{gathered}
\label{eq:22}
\end{gather}

\noindent define the respective rotation axis. Acting on the corresponding
qubits they yield

\begin{align}
\begin{split}P_{1}\left|\chi_{\pm}\right\rangle  & =\mp e^{\pm i\varphi}\left|\chi_{\mp}\right\rangle ,\\
P_{2}\left|\eta_{\pm}\right\rangle  & =i\left|\eta_{\mp}\right\rangle ,
\end{split}
\label{eq:25}
\end{align}

\noindent that is just the parity transformation in Eq. (\ref{eq:18-1}).
Thus, they constitute a family of NOT/parity quantum gates implemented
by rotations. The transformation phases can be adjusted by properly
choosing the phases in Eq. (\ref{eq:3.4}), and another interesting
choice is given by the matrices 

\begin{equation}
\begin{array}{l}
P_{3}=-\left|\chi_{+}\right\rangle \left\langle \chi_{-}\right|+\left|\chi_{-}\right\rangle \left\langle \chi_{+}\right|,\\
P_{4}=-\left|\eta_{+}\right\rangle \left\langle \eta_{-}\right|+\left|\eta_{-}\right\rangle \left\langle \eta_{+}\right|,
\end{array}\label{eq:4.6}
\end{equation}

\noindent that also belong to $\text{SU}(2)$ and which realize the
phase transformation in Eq. (\ref{eq:3.1})

\begin{equation}
\begin{split}P_{3}\left|\chi_{\pm}\right\rangle =\pm\left|\chi_{\mp}\right\rangle ,\\
P_{4}\left|\eta_{\pm}\right\rangle =\pm\left|\eta_{\mp}\right\rangle .
\end{split}
\label{eq:4.7}
\end{equation}

\noindent As a final example, choosing the trivial phases $\delta_{1}=\delta_{2}=1$
in Eq. (\ref{eq:3.4}) we obtain the matrices 

\begin{equation}
\begin{array}{l}
\tilde{P}_{1}=\left|\chi_{+}\right\rangle \left\langle \chi_{-}\right|+\left|\chi_{-}\right\rangle \left\langle \chi_{+}\right|,\\
\tilde{P}_{2}=\left|\eta_{+}\right\rangle \left\langle \eta_{-}\right|+\left|\eta_{-}\right\rangle \left\langle \eta_{+}\right|,
\end{array}\label{eq:4.8}
\end{equation}

\noindent that are still unitary, but now $\det\tilde{P}_{1}=\det\tilde{P}_{2}=-1$,
so they do not correspond to pure rotations. Their action on the qubits
is as expected

\begin{gather}
\begin{gathered}\tilde{P}_{1}\left|\chi_{\pm}\right\rangle =\left|\chi_{\mp}\right\rangle ,\\
\tilde{P}_{2}\left|\eta_{\pm}\right\rangle =\left|\eta_{\mp}\right\rangle .
\end{gathered}
\label{eq:27}
\end{gather}

\noindent Besides being unitary, the matrices $P_{1}$ through $P_{4}$
are anti-Hermitian, while $\tilde{P}_{1}$ and $\tilde{P}_{2}$ are
Hermitian, so we get the relations

\begin{equation}
\begin{array}{ll}
P_{i}^{\dagger}=P_{i}^{-1}=-P_{i} & \text{for}\, i=1,\ldots4,\\
\tilde{P}_{i}^{\dagger}=\tilde{P}_{i}^{-1}=\tilde{P}_{i} & \text{for}\, i=1,2.
\end{array}\label{eq:32-2}
\end{equation}

\noindent It should also be noticed that although $P_{1}$ and $P_{2}$
effectively realize a parity transformation, they do not have the
same effect as $\mathcal{P}^{2}$ for the states $\chi_{\pm}$. In
fact, all four matrices from $P_{1}$ to $P_{4}$ square to $-1$,
which confirm their rotational character, since it is a well known
fact that spinors change sign under a full rotation that returns them
to the starting point.

Because of Eq. (\ref{eq:18}), $P_{1}$ can also act on the states
$\left|\eta_{\pm}\right\rangle $, yielding

\begin{equation}
P_{1}\left|\eta_{\pm}\right\rangle =\mp\left|\eta_{\mp}\right\rangle .\label{eq:4.11}
\end{equation}

\noindent As for the computational basis we have

\begin{align}
\begin{gathered}P_{1}\left|0\right\rangle =-e^{i\varphi}\left|1\right\rangle ,\\
P_{1}\left|1\right\rangle =e^{-i\varphi}\left|0\right\rangle .
\end{gathered}
\label{eq:4.12}
\end{align}

\noindent Collecting the results in Eqs. (\ref{eq:25}), (\ref{eq:4.11}),
and (\ref{eq:4.12}), summarized in Table I, we see that $P_{1}$
constitutes a one parameter family of transformations with the property,
not shared with any of the other transformations, of producing the
orthogonal state up to a phase to any given pure state, including
the computational basis, and in this sense it can be regarded as a
universal quantum NOT gate.

\begin{table}

\noindent \begin{centering}
\begin{tabular}{|c|c|c|c|c|c|c|}
\hline 
$P_{1}$ & $\left|\chi_{+}\right\rangle $ & $\left|\chi_{-}\right\rangle $ & $\left|\eta_{+}\right\rangle $ & $\left|\eta_{-}\right\rangle $ & $\left|0\right\rangle $ & $\left|1\right\rangle $\tabularnewline
\hline 
 & $-e^{i\varphi}\left|\chi_{-}\right\rangle $ & $e^{-i\varphi}\left|\chi_{+}\right\rangle $ & $-\left|\eta_{-}\right\rangle $ & $\left|\eta_{+}\right\rangle $ & $-e^{i\varphi}\left|1\right\rangle $ & $e^{-i\varphi}\left|0\right\rangle $\tabularnewline
\hline 
\end{tabular}
\par\end{centering}

\protect\caption{Action of the $P_{1}$ matrix on the two types of qubits/Weyl spinors
and the computational basis.}

\end{table}

\subsection{Parity rotations}

Denoting the $\text{SU}(2)$ matrices $P_{1}$ through $P_{4}$ collectively
as $P$, it is straightforward to verify that

\begin{equation}
P\boldsymbol{\sigma}\cdot\mathbf{\hat{p}}P^{\dagger}=-\boldsymbol{\sigma}\cdot\mathbf{\hat{p}},\label{eq:34}
\end{equation}

\noindent and by the well known relation between the groups $\text{SU}(2)$
and $\text{SO}(3)$\cite{ryder1996quantum,tung1985group}, we expect
the left-hand side of Eq. (\ref{eq:34}) to induce a rotation on the
three-vector $\mathbf{p}$. Indeed, using the mapping\cite{cornwell1997group} 

\noindent 
\begin{equation}
R\left(P\right)_{ij}=\frac{1}{2}\text{Tr}\left(\sigma_{i}P\sigma_{j}P^{\dagger}\right),\label{eq:35}
\end{equation}

\noindent the induced $\text{SO}(3)$ rotation can be worked out for
all the $P$ matrices. E.g., for $P_{1}$ we obtain

\begin{equation}
R\left(P_{1}\right)=\begin{pmatrix}-\cos2\varphi & -\sin2\varphi & 0\\
-\sin2\varphi & \cos2\varphi & 0\\
0 & 0 & -1
\end{pmatrix},\label{eq:39}
\end{equation}

\noindent which is orthogonal and of unit determinant, and so it belongs
to $\text{SO}(3)$. Its action on $\mathbf{p}$ given by Eq. (\ref{eq:13}),
expressed as a column vector, yields

\begin{equation}
R\left(P_{1}\right)\mathbf{p}=-\mathbf{p}.\label{eq:40}
\end{equation}

\noindent The same result holds for the rest of the $P$ matrices
induced rotations, so we can generally write

\noindent 
\begin{equation}
R\left(P\right)\mathbf{p}=-\mathbf{p}.\label{eq:40-1}
\end{equation}

\noindent Thus, these matrices effectively provide a parity transformation
by means of a rotation, a result which is most unexpected since the
standard parity transformation in three-dimensional Euclidean space
is provided by $\mathcal{P}=-I_{3}$, minus the three-dimensional
identity matrix, and is a discrete transformation, not continuously
connected to the identity, and hence a member of $\text{O}(3)$ but
not of $\text{SO}(3)$. What, then, is the difference between these
two types of parity transformations? It is clear from their construction
that the $P$ matrices are given in the same coordinate system that
the spinors/qubits they act upon. This is also true for the induced
$R\left(P\right)$ rotations acting on three-vectors. In fact, it
can be verified that substituting the inverse mapping

\begin{gather}
\begin{gathered}\theta=\arccos\left(\frac{z}{\sqrt{x^{2}+y^{2}+z^{2}}}\right),\\
\varphi=\arctan\left(\frac{y}{x}\right)
\end{gathered}
\label{eq:40-2}
\end{gather}

\noindent in the $R\left(P\right)$ matrices, and applying them to
the column coordinate vector $\mathbf{x}=\left(x,y,z\right)$, results
in $-\mathbf{x}$. This is in contrast with the standard $-I_{3}$
parity transformation, which produces $-\mathbf{x}$ \emph{independently}
of the coordinate system used to express $\mathbf{x}$. As for the
$\tilde{P}$ matrices, they also satisfy Eq. (\ref{eq:34}), but they
belong to $\text{U}(2)$ and cannot induce an $\text{SO}(3)$ transformation,
nor can they be pure rotations.

If the Bloch vector is not the three-momentum, but instead a general
vector $\hat{\mathbf{n}}$ with the same coordinates as in Eq. (\ref{eq:13}),
all the results presented thus far still hold, with the proviso that
in this case the qubits/spinors are no longer helicity eigenstates,
but rather fixed-axis, non-relativistic spinors\cite{Dreiner:2008tw,sakurai2014modern},
and therefore not solutions to the Weyl equations. This also shows
that indeed the $P$ matrices can be regarded as quantum negation
gates, independently of the character of the Bloch vector.

\subsection{Dynamical decoupling}

Let us consider a qubit interacting with the environment (the bath),
with the general interacting Hamiltonian

\begin{equation}
\mathcal{H}_{SB}=\boldsymbol{\sigma}\cdot\mathbf{B},\label{eq:4.19}
\end{equation}

\noindent where $\mathbf{B}=\left(B_{1},B_{2},B_{3}\right)$ is the
bath vector operator. The evolution of the qubit through the bath
(free evolution) is then given by

\begin{equation}
U(t)=\exp\left(-i\mathcal{H}_{SB}t\right),\label{eq:4.20}
\end{equation}

\noindent If $\mathbf{B}$ is given in the same coordinates as $\mathbf{p}$
in Eq. (\ref{eq:13})

\begin{equation}
\mathbf{B}=\left|\mathbf{B}\right|\left(\sin\theta\cos\varphi,\sin\theta\sin\varphi,\cos\theta\right),\label{eq:4.21}
\end{equation}

\noindent then it follows from Eq. (\ref{eq:34}) that

\begin{equation}
\left\{ P,\mathcal{H}_{SB}\right\} =0,\label{eq:4.22}
\end{equation}

\noindent where the curly brackets indicate anti-commutation. Let
us now consider the following cycle: free evolution for a time $\tau$
followed by an application of a $P^{\dagger}$ transformation on the
qubit, followed by free evolution for another interval of time $\tau$,
followed by an application of a $P$ transformation on the qubit.
Using the relation $A\exp(iB)A^{\dagger}=\exp\left(iABA^{\dagger}\right)$
and Eq. (\ref{eq:4.22}) the evolution through the cycle is

\begin{equation}
P\exp\left(-i\mathcal{H}_{SB}t\right)P^{\dagger}=I_{2},\label{eq:4.23}
\end{equation}

\noindent meaning the system is perfectly decoupled from the bath.
The action of the $P$ transformations is to be understood here as
the application of pulses on the system. They can be realized as a
magnetic field applied in the direction of the given rotation axis,
e.g. the vector $\mathbf{a}$ in Eq. (\ref{eq:23}), provided the
magnetic field vector is properly normalized to preserve the unitarity
of $P$. These pulses are localized because of the coordinate dependence
of the $P$ transformations, and can be made almost instantaneous
by a sufficiently strong magnetic field.

\section{Weyl equations and parity rotations}

To simplify the notation, let the spinors, either $\chi_{\pm}(\mathbf{p})$
or $\eta_{\pm}(\mathbf{p})$, be generally denoted by $\xi_{\pm}(\mathbf{p})$.
When acted upon by the $P$ matrices, the phases in going from left-handed
spinors to right-handed ones will be denoted by $\lambda(\mathbf{p})$,
and those in going from right-handed spinors to left-handed ones by
$\lambda'(\mathbf{p})$. Thus,

\begin{gather}
\begin{gathered}P\xi_{-}=\lambda(\mathbf{p})\xi_{+},\\
P\xi_{+}=\lambda'(\mathbf{p})\xi_{-},
\end{gathered}
\label{eq:33}
\end{gather}

\noindent and Eq. (\ref{eq:12}) reads

\begin{equation}
\boldsymbol{\sigma}\cdot\mathbf{\hat{p}}\,\xi_{+}(\mathbf{p})=\xi_{+}(\mathbf{p}),\label{eq:43}
\end{equation}

\noindent for a positive energy solution to Eq. (\ref{eq:9}). The
same equation is satisfied by the negative energy solution $\xi_{+}(\mathbf{p})\exp\left\{ i\left(Et+\mathbf{x}\cdot\mathbf{p}\right)\right\} $\cite{pokorski2000gauge}.
Similarly, solutions to Eq. (\ref{eq:10}) are given by

\begin{equation}
\boldsymbol{\sigma}\cdot\mathbf{\hat{p}}\,\xi_{-}(\mathbf{p})=-\xi_{-}(\mathbf{p}).\label{eq:44}
\end{equation}

\noindent From Eq. (\ref{eq:43}) we have

\begin{equation}
P\boldsymbol{\sigma}\cdot\mathbf{\hat{p}}P^{\dagger}\, P\xi_{+}(\mathbf{p})=P\xi_{+}(\mathbf{p}),\label{eq:45}
\end{equation}

\noindent and upon using Eqs. (\ref{eq:33}) and (\ref{eq:34}) we
get back Eq. (\ref{eq:44}). In this manner we obtain a \emph{unitary}
relation between the right and left-handed Weyl equations. This is
to be contrasted with Wigner's anti-unitary case, with $P\xi_{\pm}(\mathbf{p})$
replacing $\sigma_{2}\xi_{\pm}^{*}(\mathbf{p})$, and Eq. (\ref{eq:34})
replacing $\sigma_{2}\left(\boldsymbol{\sigma}\cdot\mathbf{\hat{p}}\right)^{*}\sigma_{2}=-\boldsymbol{\sigma}\cdot\mathbf{\hat{p}}$.

In four dimensional space-time the $P$ matrices do not induce a transformation
on four-vectors. This is because the relation between $\text{SL}\left(2,C\right)$,
the set of $2\times2$ complex matrices with unit determinant, and
the restricted Lorentz group $L_{+}^{\uparrow}$\cite{tung1985group,s1995group},
consisting of rotations and Lorentz boosts, breaks down for the massless
case in consideration. This relation is established through the identification
of the four-momentum vector $p^{\mu}=(E,\mathbf{p})$ with the Hermitian
matrix

\begin{equation}
\sigma\cdot p=\begin{pmatrix}E+\left|\mathbf{p}\right|\cos\theta & e^{-i\varphi}\left|\mathbf{p}\right|\sin\theta\\
e^{i\varphi}\left|\mathbf{p}\right|\sin\theta & E-\left|\mathbf{p}\right|\cos\theta
\end{pmatrix},\label{eq:41}
\end{equation}

\noindent where $\sigma\cdot p\equiv\sigma^{\mu}p_{\mu}$, $\sigma^{\mu}=\left(I_{2},\boldsymbol{\sigma}\right)$,
and $\mathbf{p}$ is given by Eq. (\ref{eq:13}). Then the similarity
transformation $A\sigma\cdot pA^{\dagger}$, with $A\in\text{SL}\left(2,C\right)$
corresponds to the transformation $\Lambda p$ with $\Lambda\in L_{+}^{\uparrow}$.
But $\det\sigma\cdot p=\det A\sigma\cdot pA^{\dagger}=p^{2}=0$ in
the massless case, and this would induce a non-invertible transformation
on four-vectors, which cannot be a Lorentz transformation, whether
it belongs to the restricted group or not. 

Equation (\ref{eq:41}) also appears in the definition of twistors\cite{srednicki2007quantum}.
Changing the spinors normalization from unity to $\sqrt{2E}=\sqrt{2\left|\mathbf{p}\right|}$
we readily obtain the defining relation

\begin{equation}
\sigma\cdot p=2\left|\mathbf{p}\right|\xi_{+}\xi_{+}^{\dagger},\label{eq:41-1}
\end{equation}

\noindent for the twistors $\sqrt{2\left|\mathbf{p}\right|}\xi_{+}$
and $\sqrt{2\left|\mathbf{p}\right|}\xi_{+}^{\dagger}$, which are
just positive helicity spinors/qubits and their Hermitian conjugates.
On the other hand, transforming $\sigma\cdot p$ with the $P$ matrices
gives an alternative definition in terms of $\sqrt{2\left|\mathbf{p}\right|}\xi_{-}$
and $\sqrt{2\left|\mathbf{p}\right|}\xi_{-}^{\dagger}$

\begin{equation}
P\sigma\cdot pP^{\dagger}=\begin{pmatrix}E-\left|\mathbf{p}\right|\cos\theta & -e^{-i\varphi}\left|\mathbf{p}\right|\sin\theta\\
-e^{i\varphi}\left|\mathbf{p}\right|\sin\theta & E+\left|\mathbf{p}\right|\cos\theta
\end{pmatrix}=2\left|\mathbf{p}\right|\xi_{-}\xi_{-}^{\dagger},\label{eq:42}
\end{equation}

\noindent whose spatial part is just Eq. (\ref{eq:34}).

\section{Concluding remarks}

In this letter I have shown a new class of transformations with the
remarkable property of effecting a space inversion while being pure
rotations, and use them to realize an operation thought not to be
possible before, namely, the fact that orthogonal pure state qubits
can be unitarily transformed into each other, and the same for Weyl
spinors of different helicity. For qubits, the transformations provide
a new kit of parity gates, and in one instance they act as universal
NOT gates. They also serve the purpose of canceling the coupling with
the environment in dynamical decoupling schemes. For Weyl spinors
they provide a new symmetry transformation for the Weyl equations.
I have also shown an equivalence between qubits and Weyl spinors which
could foster interdisciplinary research. 

The transformations here presented require a complete characterization
within the theory of Lie groups, and by themselves constitute a relevant
result in mathematical physics, but they could also find useful applications
in other areas where this theory is relevant, such as condensed matter
and high-energy physics. There is also the possibility to apply them
in quantum optics, and to use them to improve other areas of quantum
information, such as the fidelity of quantum anti-cloning.

\bibliographystyle{apsrev4-1}

\end{document}